\documentclass[runningheads]{llncs}

\usepackage{graphicx}
\usepackage{xcolor}
\usepackage{multirow}
\begin{document}
\title{Accelerating RNN-based Speech Enhancement on a Multi-Core MCU with Mixed FP16-INT8 Post-Training Quantization %\thanks{Supported by organization x.}
}
\titlerunning{Accelerating RNN-based SE on an MCU with Mixed FP16-INT8 PTQ}
\author{Manuele Rusci \inst{1,2} \and % \orcidID{0000-1111-2222-3333}
Marco Fariselli \inst{2} \and % \orcidID{0000-1111-2222-3333}
Martin Croome \inst{2} \and \\ % \orcidID{0000-1111-2222-3333}
Francesco Paci \inst{2} \and % \orcidID{0000-1111-2222-3333}
Eric Flamand \inst{2} % \orcidID{0000-1111-2222-3333}
}
\authorrunning{M. Rusci et al.}
\institute{Universita' di Bologna, Bologna, ITA 
\\ \email{manuele.rusci@unibo.it} \and
Greenwaves Technologies, Grenoble, FRA\\
\email{\{name.surname\}@greenwaves-technologies.com}}
\maketitle              % typeset the header of the contribution

\begin{abstract}
{ 
This paper presents an optimized methodology to design and deploy Speech Enhancement (SE) algorithms based on Recurrent Neural Networks (RNNs) on a state-of-the-art MicroController Unit (MCU), with 1+8 general-purpose RISC-V cores. To achieve low-latency execution, we propose an optimized software pipeline interleaving parallel computation of LSTM or GRU recurrent blocks, featuring vectorized 8-bit integer (INT8) and 16-bit floating-point (FP16) compute units, with manually-managed memory transfers of model parameters. 
To ensure minimal accuracy degradation with respect to the full-precision models,
we propose a novel FP16-INT8 Mixed-Precision Post-Training Quantization (PTQ) scheme that compresses the recurrent layers to 8-bit while the bit precision of remaining layers is kept to FP16.
Experiments are conducted on multiple LSTM and GRU based SE models trained on the Valentini dataset, featuring up to 1.24M parameters.
Thanks to the proposed approaches, 
we speed-up the computation by up to 4$\times$ with respect to the lossless FP16 baselines. Differently from a uniform  8-bit quantization that degrades the PESQ score by 0.3 on average, the Mixed-Precision PTQ scheme leads to a low-degradation of only 0.06, while achieving a 1.4-1.7$\times$ memory saving. Thanks to this compression, we cut the power cost of the external memory by fitting the large models on the limited on-chip non-volatile memory and we gain a MCU power saving of up to 2.5$\times$ by reducing the supply voltage from 0.8V to 0.65V while still matching the real-time constraints. 
Our design results $>$10$\times$ more energy efficient than state-of-the-art SE solutions deployed on single-core MCUs that make use of smaller models and quantization-aware training. 
}


\end{abstract}

\section{Introduction}
Novel speech-centric devices, e.g. miniaturized Hearing Aids, make use of AI-based methods to process audio data in real-time for improving the signal intelligibility. 
Given the small sizes, these devices present a limited energy budget: a lifetime of up to 20h can be achieved with a small 60mAh battery if the average power consumption is 10mW, considering sensing, computation and actuation costs. 
Because of the severe energy constraints, low-power Micro-Controller Units (MCUs) are typically chosen as Digital Processing Units to handle control and processing tasks. These processing units feature a limited computational power (single core CPU) 
and up to few MB of on-chip memory, making the integration process of complex AI speech processing pipelines extremely challenging.

Speech Enhancement (SE), the ability of removing background noises from a (noisy) audio signal, is getting popular among the AI capabilities of speech sensors. While in the past SE methods relied on digital signal processing filters~\cite{boll1979suppression,cohen2001speech}, recent approaches integrate Deep Learning (DL) strategies, which have demonstrated a superior effectiveness to deal with non-stationary noises~\cite{reddy2019scalable}. 
To cancel out noise components, DL based approaches learn in a supervised fashion to estimate spectral suppression masks from a set of features extracted from the noisy speech. Among the causal models tailored for real-time computation, Recurrent Neural Networks have shown promising results~\cite{xia2020weighted,hu2020dccrn,valin2020perceptually,liu2022drc}. 
These approaches encode the input signal, typically in the frequency domain (e.g. STFT or Mel spectrograms), into an embedding vector that feeds one or multiple recurrent layers, i.e. GRU or LSTM, acting also as memory components of the RNN based SE filter. 
The cleaned audio signal is reconstructed by decoding the outputs of the recurrent layers in a frame-by-frame streaming fashion.

Unfortunately, current DL methods target real time execution on high-end devices~\cite{reddy2021interspeech} and are not fully-optimized for MCUs. Only~\cite{jianjia_ma_2020_4158710} and~\cite{fedorov2020tinylstms} described  design methodologies of RNN based SE models, with less than 0.5M parameters, for single-core MCUs. 
More in details, the \textit{NNoM} framework was used to deploy the RNNoise model~\cite{valin2018hybrid} on a single-core ARM Cortex-M MCU~\cite{jianjia_ma_2020_4158710}. 
The RNNoise algorithm includes small GRU layers with constrained activation ranges, leading to an effective 8-bit quantization. On the other side,  TinyLSTM~\cite{fedorov2020tinylstms} made use of Quantization-Aware Training (QAT)~\cite{jacob2018quantization} to compress an LSTM based model to 8 bit without accuracy degradation. Despite its effectiveness, the QAT technique is not always applicable because of the additional compute and data resources needed to simulate the non-linear quantization error at (re-)training time~\cite{gholami2021survey}. 
Hardware-specific fine-tuning such as Block Pruning has been also developed to efficiently map SE RNNs on MicroNPU accelerators~\cite{stamenovic2021weight}
Differently from these solutions, \textit{(i)} we aim at a lossless and low-cost Post-Training Quantization methodology for scalable RNN-based SE algorithms and \textit{(ii)} we investigate an optimized deployment flow for general-purpose multi-core MCUs, to achieve a final design more energy-efficient than state-of-the-art solutions.

To this aim, we combine multiple strategies. 
Firstly, we target a multi-core compute platform with 1+8 RISC-V CPUs, featuring 8-bit integer (INT8) and 16-bit floating-point (FP16) MAC vector units.
Secondly, we design an optimized software pipeline, in charge of scheduling at runtime parallel compute calls with manually-managed memory transfers, also from external L3 memories. 
To gain an almost lossless compression, we also propose a novel Mixed-Precision FP16-INT8 (MixFP16-INT8) Post-Training Quantization scheme, which quantizes only the RNN parameters and activations to INT8 while keeping the bit precision of other tensors to FP16.
% Contributions
This paper makes the following contributions:
\begin{itemize}
    \item We present an optimized HW/SW design for LSTM and GRU based SE models for multi-core MCU systems with limited memory space.
    \item We propose an almost lossless Mixed-Precision FP16-INT8 Post-Training Quantization scheme to accelerate RNN-based SE on MCUs.
    \item We provide a detailed analysis of latency and HW/SW efficiency on a 22-nm RISC-V 1+8-core MCU.
\end{itemize}

% results
Our work demonstrates, for the first time, an optimized design for RNN-based SE models relying only on PTQ, without any need for expensive QAT, with Mixed-Precision FP16-INT8.
When benchmarked on the Valentini dataset, the RNN trained models show an average reduction of the PESQ and STOI scores of only 0.06 and 0.007. The proposed HW/SW design results  $>$10$\times$ more energy efficient than state-of-the-art  solutions deployed on single-core MCUs.

\begin{figure}[t]
\centering
\includegraphics[width=0.9\textwidth]{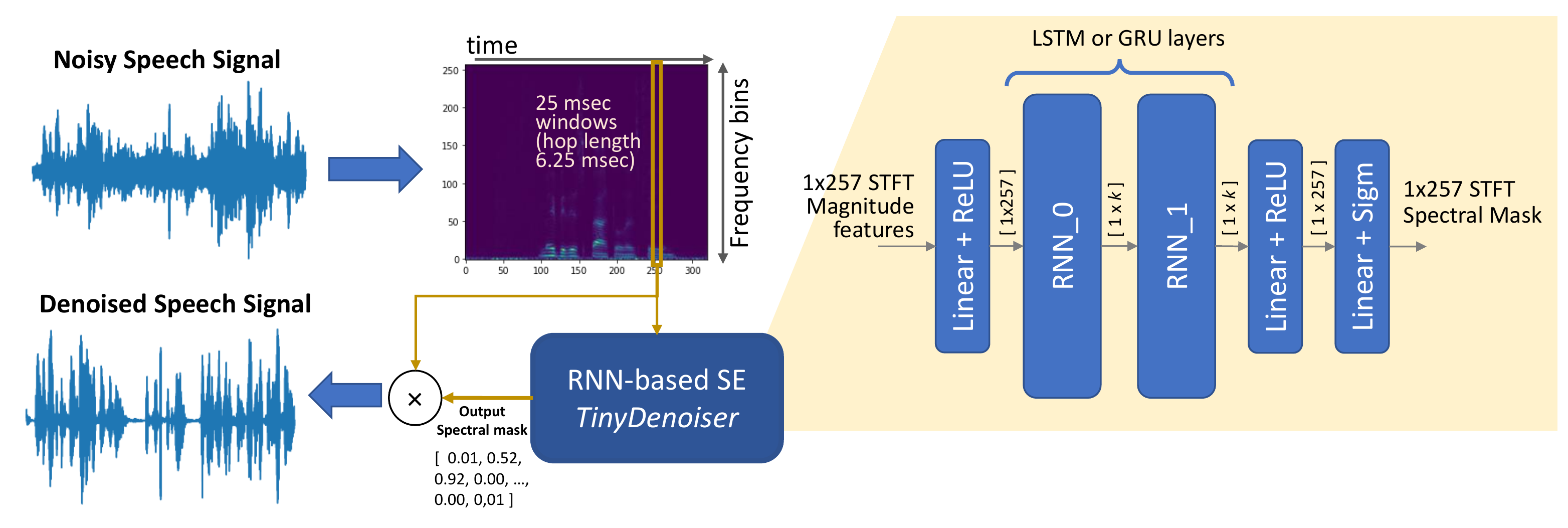}
\caption{TinyDenoiser models for Speech Enhancement on MCUs.} 
\label{fig:tinydenoiser}
\end{figure}

\begin{table}[t]

\caption{Characteristics of the RNN-based TinyDenoiser variants.}
\label{tab:tinydenoiser}
\centering
\scriptsize

\begin{tabular}{l|cccc}
\textbf{}              & \textbf{LSTM256} & \textbf{GRU256} & \textbf{LSTM128} & \textbf{GRU128} \\ \hline
\textit{\textbf{k}}             & 256              & 256             & 128              & 128             \\
\textbf{RNN\_0 layer}        & LSTM(257,256)    & GRU(257, 256)   & LSTM(257,128)    & GRU(257, 128)   \\
\textbf{RNN\_1 layer}        & LSTM(257,256)    & GRU(257, 256)   & LSTM(128 128)    & GRU(128, 128)   \\
\textbf{Params}        & 1.24 M           & 0.985 M         & 0.493 M          & 0.411 M         \\
\textbf{\% rnn params} & 84\%             & 80\%            & 66.50\%          & 59.80\%        
\end{tabular}

\end{table}

\section{RNN based Speech Enhancement on Multi-Core MCUs}
This Section firstly describes the scalable RNN-based SE model family, denoted as \textit{TinyDenoiser}, that we consider for this study. Second, we detail
the target HW platform and the mapping of the  %the memory allocation strategy on the target HW platform and we describe the inference execution model based on the 
proposed software pipeline. 
Lastly, we present our novel Mixed-Precision FP16-INT8 PTQ method.
%to achieve low-energy computation and low accuracy degradation with respect to the full-precision models. 

\subsection{TinyDenoiser models}
\label{sec:tinydenoiser}
Fig.~\ref{fig:tinydenoiser} shows the \textit{TinyDenoiser} pipeline. The model takes as input the STFT frequency map of a noisy speech signal and predicts a spectral gain mask, whose values are in the range $[0,1]$.
In more detail, the audio input is sampled at 16kHz and the STFT frequency features are computed over a 25 msec audio frame, after Hanning windowing.
For every audio frame, a total of 257 STFT magnitude values are fed into the model and 257 gain values are returned as output.
The hop size is 6.25 msec (25\% of the window length), which determines the real-time constraint of the inference task.
The filtered frequency spectrum of the audio frame, computed by masking the noisy spectrum, is converted back to the time domain using an inverse STFT transform. In a real-time streaming processing scenario, the denoised speech signal is obtained by overlap-and-add operations of the cleaned audio frames.

Drawing inspiration from TinyLSTM~\cite{fedorov2020tinylstms}, the TinyDenoiser includes two RNN layers with a parametric output size of length $k$, a Fully-Connected (FC) input layer producing 257 features and two final FC layers both producing 257 features. 
With the exception of the last layer, which features a Sigmoid activation to estimate the frequency gains, the other FC layers are followed by a batchnorm and a ReLU activation. Concerning the RNN layers, we experiment with both LSTM and GRU layers with an output size of $k = \{128,256\}$, originating multiple variants of the TinyDenoiser denoted as \textit{LSTM256}, \textit{GRU256}, \textit{LSTM128} and \textit{GRU128}. As reported in Table~\ref{tab:tinydenoiser}, these variants feature a number of parameters ranging from 0.4M and 1.24M. Note that the majority of the parameters (and the operations) are due to the RNN layers (up to 84\% for \textit{LSTM256}).

\begin{figure}[t]
\centering
\includegraphics[width=0.9\textwidth]{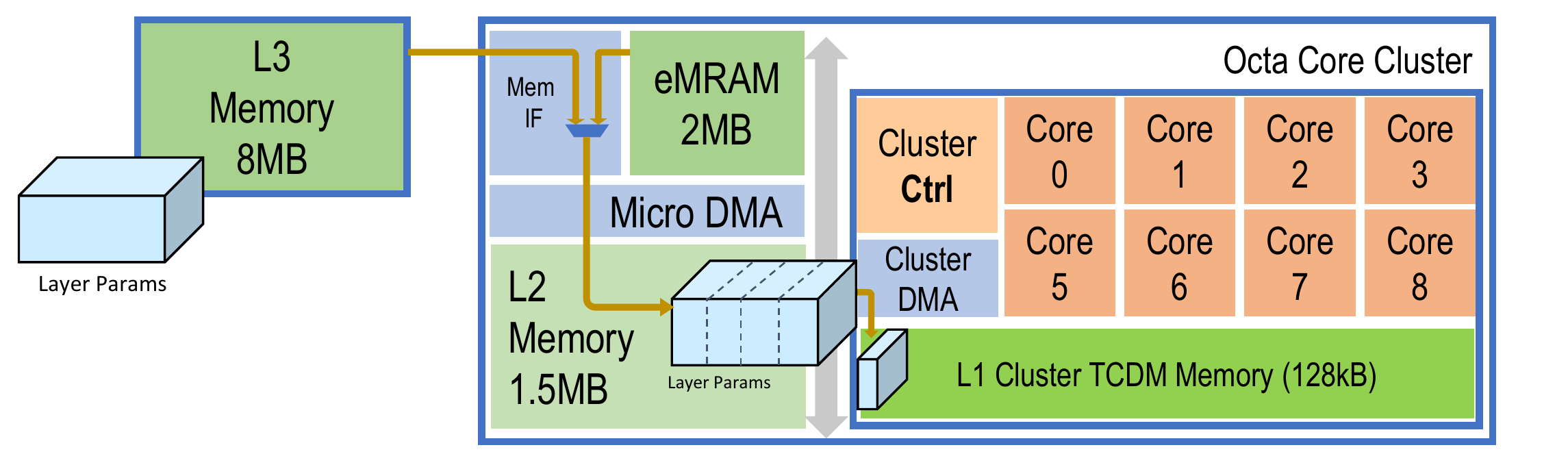}
\caption{Micro-Architecture of the target platform. The cluster on the right includes 1+8 cores. An external memory or an on-chip non-volatile memory can  be used to permanently store the model parameters. } 
\label{fig:gap-rnn}
\end{figure}

% Tiny denoiser model
\subsection{Memory Management for RNN deployment on the target HW} \label{sec:HWtarget}
Fig.~\ref{fig:gap-rnn} depicts the architecture of the MCU platform targeted for the deployment of the RNN-based SE model. 
Internally, the system includes a cluster with 8 RISC-V CPUs tailored for computation and 1 core for control operation, denoted as the Cluster Controller (CC). The 1+8 cores can load data from a 128kB Tightly Coupled Data Memory, namely the L1 memory,  in a single clock-cycle. Note that the L1 memory is not a data cache.
Every core has a 8-bit MAC vector unit, capable of computing a dot-product between two $4\times$8-bit vectors and accumulation in a single clock-cycle (i.e. 4 MAC/clk), while 4 floating point units are shared among the 8 compute cores, implementing single-cycle  $2\times$FP16 vector MAC (2 MAC/clk). 
Outside the cluster, the platform includes a 1.5 MB L2 memory; the cluster cores can access data from the L2 memory $\sim 10 \mathrm{x}$ slower than accessing the L1 memory.
%with  a $\sim 10 \mathrm{x}$ higher access latency than the L1 memory. 
To reduce this overhead, the cluster DMA can be programmed by the CC core to copy data between L2 and L1 memories with a peak bandwidth of $8 \mathrm{Byte/clk}$.
In the background of the DMA operations, the Control Core dispatches and synchronizes parallel tasks on the 8 compute cores.

To deploy the RNN-based TinyDenoiser on the HW target platform, the layer parameters are permanently stored into a non-volatile memory. Because the storage requirements can grow up to several MBs, we use an off-chip FLASH memory (\textit{ExtFLASH}), also denoted as L3 memory, with a capacity of 8MB and connected to the MCU via OctoSPI. Data can be copied from L3 to L2 memories in the background of other operations by programming the MicroDMA module. Note that the IO memory interface reaches a max bandwidth of $1 \mathrm{Byte/clk}$, 8$\times$ slower than the L2 peak bandwidth.
Alternatively, the on-chip \textit{eMRAM} non-volatile memory can be used for permanent storage, gaining a lower power consumption and a higher bandwidth but the total capacity reduces to 2MB. 

At runtime, but before entering the infinite inference loop, layer-wise network parameters can be copied from L3 (either \textit{ExtFLASH} or \textit{eMRAM}) to L2, based on the available space. 
Thanks to this process, named \textit{tensor promotion}, 
%the average latency of parameters into the L1 memory 
the time to copy parameters to the L1 memory during inference
decreases linearly with respect to the amount of promoted tensors.
If a parameter tensor does not fit the available L2 parameter buffer space, it is partitioned in sub-tensors that are sequentially loaded from L3 to L2. 
Besides storing the promoted parameters, the L2 memory must reserve space to store an activation buffer, for temporarily keeping the activation feature maps, and a parameter buffer, serving the dynamic load of not-promoted parameters from L3 to L2. 

During the inference task, the L1 memory inside the cluster acts as the working memory because of the fast access time from the compute cores: parameters and activation features are copied to this memory and then fetched concurrently by the cores. 
Because of the small size, the L1 memory is kept free from static tensor allocation.
Activation or parameter tensors or sub-tensors are rather loaded from L2 to L1 at inference time using the Cluster DMA, as depicted in Fig.~\ref{fig:gap-rnn}.

\begin{figure}[t]
\centering
\includegraphics[width=\textwidth]{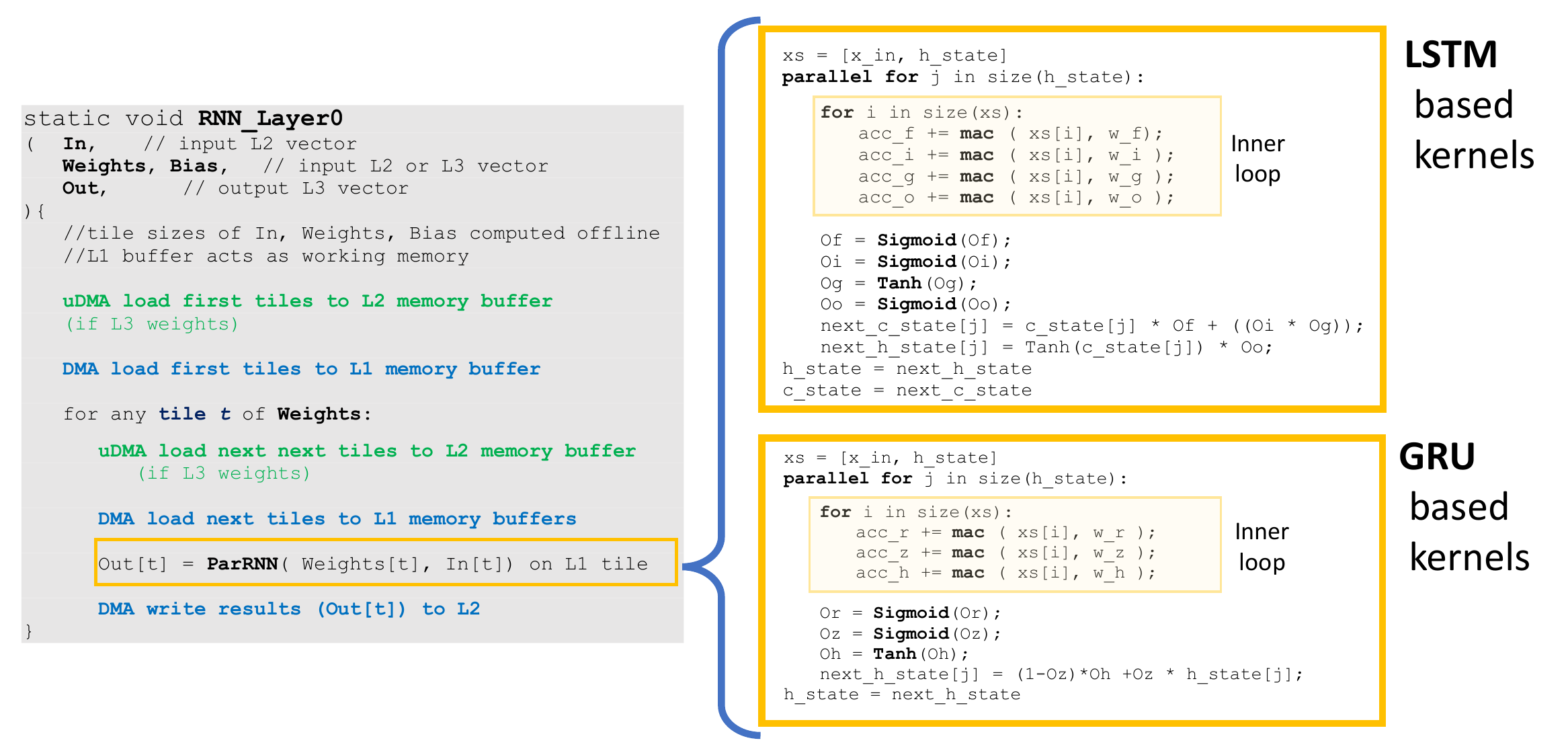}
\caption{Pseudo C code of layer-wise RNN processing. The CC core runs the code on the left. The parallel GRU and LSTM basic kernels dispatched on the compute cores are on the right. Biases are omitted for simplicity. } 
\label{fig:rnn-compute}
\end{figure}

\subsection{SW Computation Model}
\label{sec:swarchi}

The CC core runs the RNN-based SE inference SW code, which includes a sequence of layer-wise processing calls. Fig.~\ref{fig:rnn-compute} shows the pseudo C-code for a RNN layer processing task; the same software architecture applies for FC layer processing.
The input and output activation tensor arguments, including the RNN states, are L2 memory arrays. 
On the contrary, the RNN parameter array (\textit{Weights}) can be stored in L2 or L3 memory, depending if any promotion occurred as discussed before. 

Every layer-wise function interleaves data copies from L3 and L2 memories to the L1 memory and calls to the compute tasks. These latter are dispatched and parallelized over the 8 compute cores of the cluster. 
To be more specific, the CC core programs the MicroDMA and Cluster DMA modules to operate, respectively, asynchronous data copies from L3 to L2 and from L2 to L1. Note that L3 transfers occurs only if layer parameters are not promoted to L2; in this case the MicroDMA is not programmed. 

Typically, input, weight and output tensors of a layer cannot entirely fit the L1 memory (limited to 128kB). For this reason, large tensors are sliced in sub-tensors, also referred as tiles, during the code generation process. The size of the tiles are computed such as to maximize the memory occupation of the available L1 memory. Therefore the layer-wise software routine implements a \texttt{for} loop to sequentially loads (with the DMAs) the tensor slices in the background of the computation that applies on the previously copied data (Fig.~\ref{fig:rnn-compute} on the left). To realize this mechanism, we double the memory requirement of the tensor slices to account both the L1 memory needed by the compute cores and the memory buffer used by the DMA.

Based on the proposed execution model, the minimal theoretical latency $\tilde{t}_{layer}$ to process a layer can be estimated as:
\begin{equation} \label{eq:latency}
    \tilde{t}_{layer} = N_{tiles} \cdot \mathrm{max}(t_{dma}^{L3-L2}, t_{dma}^{L2-L1}, t_{core})
\end{equation}

where $N_{tiles}$ is the number of tiles, $t_{dma}^{L2-L1}$ and $t_{dma}^{L3-L2}$ are the latencies required by, respectively, the Cluster DMA and the MicroDMA to copy a single data tile from L2 to L1 and L3 to L2 and $t_{core}$ is the compute time due to the parallel task. 
Based on HW architecture described in Section~\ref{sec:HWtarget}, $  t_{dma}^{L3-L2} \approx 8 \times t_{dma}^{L2-L1} $ if considering an external SPI flash. $t_{dma}^{L3-L2}$ decreases when using instead the on-chip non-volatile memory (up to 2.6$\times$ for eMRAM).

%On the other side, 
Fig.~\ref{fig:rnn-compute} shows on the right more in details the parallel SW kernels for LSTM and GRU computation. Both kernels consists of a 2 nested loops. The outer loop, which is parallelized over the available compute cores, iterates over the size of the output feature tile. 
The inner loop computes the MAC between the combination of input features and the previous state and the GRU or LSTM weight tensors. More specifically, we target INT8 or FP16 computation depending on the used quantization type. 
To speed-up the computation of the dot-product, we exploit vectorized INT8 and FP16 MAC instructions, which can perform respectively 4 or 2 MAC/cyc per core.
%The output and the current state of the values are computed from the result of the dot-product. 
Concerning the FP16 \texttt{tanh} and \texttt{sigmoid} functions applied on the accumulators, we use a fast approximation exploiting vectorized FP16 instructions, while the INT8 version makes use of LUTs. 

Because of the high number of iterations of the inner loop, the total latency of the kernel is typically dominated by the computation of this loop. For INT8 LSTM and GRU, we account a minimal theoretical per-core latency of 9 (5 vect LD + 4 vect MAC) and  7 (4 vect LD + 3 vect MAC) clock cycles to compute 4$\times$4 and 3$\times$4 MAC operations, respectively. 
In case of FP16, the software kernel computes half of the MAC operations during the same period, if not considering the stalls occurring while accessing concurrently the shared floating-point units.

Note that the peak computation power scales linearly with number of compute cores, up to reach the memory bottleneck (see Eq.\ref{eq:latency}).
In fact, as we increase the number of compute cores, the total bandwidth requirement for RNN computation exceeds the capacity of the target platform for both L3 and L2 memories. For instance, a FP16 LSTM layer processing on 8 cores demands for 8 (cores) $\times$ 5 (LD) $\times$  2 (FP16 datatype) bytes every 9 cycles, which is much higher than the bandwidth from ExtFlash memory (1 byte/clk). In this case, using a lower datatype, e.g. 8-bit, results in faster computation for multiple reasons. Firstly, the memory bandwidth requirements of INT8 kernels is 2$\times$ lower than FP16 ones. Secondly, the 2$\times$ higher memory saving can lead the model parameters to entirely fit the on-chip non-volatile eMRAM memory. Lastly, a smaller tensor parameters can be promoted permanently to the L2 memory. On the other side, INT8 leads to a higher quantization error with respect to a full-precision model than FP16, potentially affecting the prediction quality of the full model. Our solution to this problem is discussed in the next section.

\subsection{Mixed FP16-INT8 Post-Training Quantization}
%To deploy the TinyDenoiser models on the target platform, we quantize model parameters and activation features to reduce the memory requirements and speed up the computation (INT8 kernels are $\sim 2 \times$ faster than FP16 kernels). 
TinyDenoiser models are quantized with Post-Training Quantization. 
We refer to the IEEE 754 standard for the FP16 format and quantization, i.e. a casting. On the other side, we follow~\cite{jacob2018quantization} for INT8 symmetric quantization. According to this, every full-precision tensor $x$ is approximated as an integer tensor $X$ as:
\begin{equation}
\label{eq:quant}
    X = \biggl \lfloor  \frac{clamp(x, q_{min}, q_{max} )}{S} \biggl \rceil , 
    \qquad S =  \frac{q_{max} - q_{min} }{2^n-1}
 %   x = \frac{q_{max} - q_{min} }{2^n-1} \cdot ( X - Z)
\end{equation}
where $n$ is the number of bits, $S$ is the scale factor, which impacts the conversion resolution, and $[q_{min}, q_{max}]$ is the quantization range of an individual tensor. 
%Note that we refer to a symmetric quantization scheme, as it is currently supported by the software because of the higher performance than an asymmetric scheme. 
%to avoid a Quantization-Aware Training procedure that is expensive either in terms of computing resources and time and subjected to the availability of the training data. 
In particular, the PTQ routine estimates the quantization range of activation tensors (Eq.~\ref{eq:quant}) by collecting the intermediate tensor statistics after feeding a trained model with a set of calibration samples. 
For the parameters, we refer to the min/max values.

For RNN-based TinyDenoiser models, we observe a degraded quality, measured using objective metrics (see Sec. \ref{sec:result}), if using a uniform 8-bit quantization on the whole model.
On the contrary, the FP16 quantization works lossless. 
We hypothesize the INT8 accuracy drop to originate from unbounded tensor ranges, e.g. STFT input or ReLU output, which are clamped after quantization (Eq.~\ref{eq:quant}) or causing a large scale factor $S$. 
Quantization error propagates also over time on RNN-based models.  
%For RNN-based models, it must also be taken into consideration that the quantization error propagates over time. 
However, we noticed that both LSTM layers and GRU layers use constrained activation functions (\texttt{tanh} and \texttt{sigmoid}), i.e. output features maps features a numerical range limited by design, with the exception of the \textit{LSTM C\_state}. This motivates us \textit{to quantize only the RNN layers, which demands the highest memory requirement of the whole model, to INT8 while leaving the rest to FP16}. We named this quantization as \textbf{Mixed-Precision FP16-INT8}, also referred in short as MixFP16-INT8.
To this aim, we restrict the tensor statistic collection during PTQ to the input, states and output values of the RNN layers. In addition, two extra-layers, computationally inexpensive, are inserted in the inference graph for data type conversion purpose between FP16 and INT8 nodes and viceversa, according to Eq.~\ref{eq:quant}. 

\section{Experimental Results}
\label{sec:result}
Before showing the effectiveness, in terms of memory, latency and energy gains, of our optimized design, we report the accuracy of the trained SE models after quantization. Lastly, we compare our approach with state-of-the-art  solutions.

% Experimental Setup

\begin{table}[t]
\caption{STOI and PESQ scores and memory footprint of the quantized TinyDenoiser models after Post-Training Quantization using FP16, INT8 or MixFP16-INT8 options.}
\centering
\scriptsize
\label{tab:accuracy}
\begin{tabular}{|l|c|c|ccc|ccc|ccc|ccc|}
\hline
\multirow{2}{*}{\textbf{Quant}}                                                   & \multirow{2}{*}{\textbf{\begin{tabular}[c]{@{}c@{}}Clamp\\ FC\end{tabular}}} & \multirow{2}{*}{\textbf{\begin{tabular}[c]{@{}c@{}}Clamp \\ RNN\end{tabular}}} & \multicolumn{3}{c|}{\textbf{LSTM256}}                                                            & \multicolumn{3}{c|}{\textbf{GRU256}}                                                             & \multicolumn{3}{c|}{\textbf{LSTM128}}                                                            & \multicolumn{3}{c|}{\textbf{GRU128}}                                                             \\ \cline{4-15} 
                                                                                  &                                                                              &                                                                                & \multicolumn{1}{c|}{\textbf{pesq}} & \multicolumn{1}{c|}{\textbf{stoi}}  & \textbf{MB}           & \multicolumn{1}{c|}{\textbf{pesq}} & \multicolumn{1}{c|}{\textbf{stoi}}  & \textbf{MB}           & \multicolumn{1}{c|}{\textbf{pesq}} & \multicolumn{1}{c|}{\textbf{stoi}}  & \textbf{MB}           & \multicolumn{1}{c|}{\textbf{pesq}} & \multicolumn{1}{c|}{\textbf{stoi}}  & \textbf{MB}           \\ \hline
\textbf{FP32}                                                                     &                                                                              &                                                                                & \textbf{2.78}                      & \multicolumn{1}{c|}{\textbf{0.942}} & 4.75                  & \textbf{2.78}                      & \multicolumn{1}{c|}{\textbf{0.942}} & 3.76                  & \textbf{2.76}                      & \multicolumn{1}{c|}{\textbf{0.941}} & 1.88                  & \textbf{2.69}                      & \multicolumn{1}{c|}{\textbf{0.940}} & 1.56                  \\ \hline
\textbf{FP16}                                                                     &                                                                              &                                                                                & \textbf{2.78}                      & \multicolumn{1}{c|}{\textbf{0.942}} & 2.37                  & \textbf{2.78}                      & \multicolumn{1}{c|}{\textbf{0.942}} & 1.88                  & \textbf{2.76}                      & \multicolumn{1}{c|}{\textbf{0.941}} & 0.94                  & \textbf{2.69}                      & \multicolumn{1}{c|}{\textbf{0.940}} & 0.78                  \\ \hline
\multirow{3}{*}{\textbf{INT8}}                                                    & \textit{max}                                                                 & \textit{max}                                                                   & \textbf{2.42}                      & \multicolumn{1}{c|}{\textbf{0.922}} & \multirow{3}{*}{1.18} & 2.19                               & \multicolumn{1}{c|}{0.932}          & \multirow{3}{*}{0.93} & 2.40                               & \multicolumn{1}{c|}{\textbf{0.922}} & \multirow{3}{*}{0.47} & 2.20                               & \multicolumn{1}{c|}{\textbf{0.925}} & \multirow{3}{*}{0.39} \\ \cline{2-5} \cline{7-8} \cline{10-11} \cline{13-14}
                                                                                  & \textit{std3}                                                                & \textit{std3}                                                                  & 2.35                               & \multicolumn{1}{c|}{0.885}          &                       & 2.17                               & \multicolumn{1}{c|}{0.902}          &                       & 2.51                               & \multicolumn{1}{c|}{0.911}          &                       & 2.12                               & \multicolumn{1}{c|}{0.892}          &                       \\ \cline{2-5} \cline{7-8} \cline{10-11} \cline{13-14}
                                                                                  & \textit{std3}                                                                & \textit{max}                                                                   & 2.34                               & \multicolumn{1}{c|}{0.886}          &                       & \textbf{2.48}                      & \multicolumn{1}{c|}{\textbf{0.929}} &                       & \textbf{2.51}                      & \multicolumn{1}{c|}{0.910}          &                       & \textbf{2.36}                      & \multicolumn{1}{c|}{0.910}          &                       \\ \hline
\multirow{2}{*}{\textbf{\begin{tabular}[c]{@{}l@{}}MixFP16\\ -INT8\end{tabular}}} & \textit{std3}                                                                & \textit{max}                                                                   & 2.69                               & \multicolumn{1}{c|}{0.926}          & \multirow{2}{*}{1.37} & \textbf{2.72}                      & \multicolumn{1}{c|}{\textbf{0.940}} & \multirow{2}{*}{1.13} & 2.67                               & \multicolumn{1}{c|}{0.925}          & \multirow{2}{*}{0.67} & \textbf{2.63}                      & \multicolumn{1}{c|}{0.935}          & \multirow{2}{*}{0.55} \\ \cline{2-5} \cline{7-8} \cline{10-11} \cline{13-14}
                                                                                  & \textit{max}                                                                 & \textit{max}                                                                   & \textbf{2.73}                      & \multicolumn{1}{c|}{\textbf{0.930}} &                       & 2.56                               & \multicolumn{1}{c|}{0.941}          &                       & \textbf{2.69}                      & \multicolumn{1}{c|}{\textbf{0.927}} &                       & 2.49                               & \multicolumn{1}{c|}{\textbf{0.937}} &                       \\ \hline
\end{tabular}
\end{table}

\subsection{Accuracy after Mixed-Precision PTQ}
We train the TinyDenoiser models on the Valentini dataset~\cite{valentini2017noisy}, which consists of clean and noisy speech audio clips from 28 speakers sampled at 16kHz.  
The training environment is taken from~\cite{defossez2020real}: the loss functions is a weighted combination of the L1 loss and the STFT loss and an ADAM optimizer is used with a learning rate of 3e-4. 
We use a batch size of 64 and set 200 epochs of training. At the end of the training procedure, we select the trained model with the highest score on a validation dataset composed by audio clips of speakers \textit{p286} and \textit{p287}, opportunely removed from the train set. 
For evaluation purpose, we refer to the PESQ and STOI objective metrics. 

We implement the Post-Training Quantization procedure as a module of the GAP\textit{flow} toolset\footnote{ \url{https://greenwaves-technologies.com/tools-and-software/}}.
The script imports a trained full-precision graph and quantizes it to FP16, INT8 or MixFP16-INT8, before generating the application code for deployment purpose. 
We use 4 randomly-chosen samples of the validation set (\textit{p286\_035}, \textit{p286\_166}, \textit{p287\_151}, \textit{p287\_269}) for the calibration of the quantization ranges. In particular, we consider either the maximum absolute values of the activation parameters $x$ or $q_{max} = \mathrm{mean}(x) + 3  \cdot \mathrm{std\_dev}(x)$, that we denote as \textit{max} and \textit{std3} calibration settings, respectively. Additionally, we make use of a moving average filter in the estimation of the quantization ranges when feeding the models with multiple calibration samples as done in~\cite{jacob2018quantization}.  

Table~\ref{tab:accuracy} reports the PESQ and STOI scores of the TinyDenoiser models on the Valentini test dataset after PTQ, together with the memory occupation (in MB) of the whole quantized parameters. The FP16 models are lossless with respect to FP32 trained models but gain $2\times$ memory saving. On the contrary, despite the additional $2\times$ memory compression factor, a uniform 8-bit quantization leads to a score degradation of, on average, 0.3 and 0.015 concerning the PESQ and the STOI metrics, respectively. 
We applied multiple combinations of \textit{max} and \textit{std3} quantization ranges to the RNN layers activations (Clamp RNN in the table) or the FC layers, including the input of the SE model. 
For INT8, we observed \textit{max} quantization ranges to bring benefits to the RNN layer quantization, therefore we applied this setting also for MixFP16-INT8 quantization. On the contrary, we have not found any experimental evidence to select between \textit{std3} or \textit{max} on other layers. 
Overall, our proposed Mixed Precision FP16-INT8 PTQ recovers the accuracy degradation of INT8: on average, PESQ and STOI scores result to degrade of only 0.06 and 0.007, respectively. The effectiveness of the approach is also assessed by the 1.4–1.7$\times$ less memory to store the model parameters.

\begin{figure}[t]
\centering
\includegraphics[width=\textwidth]{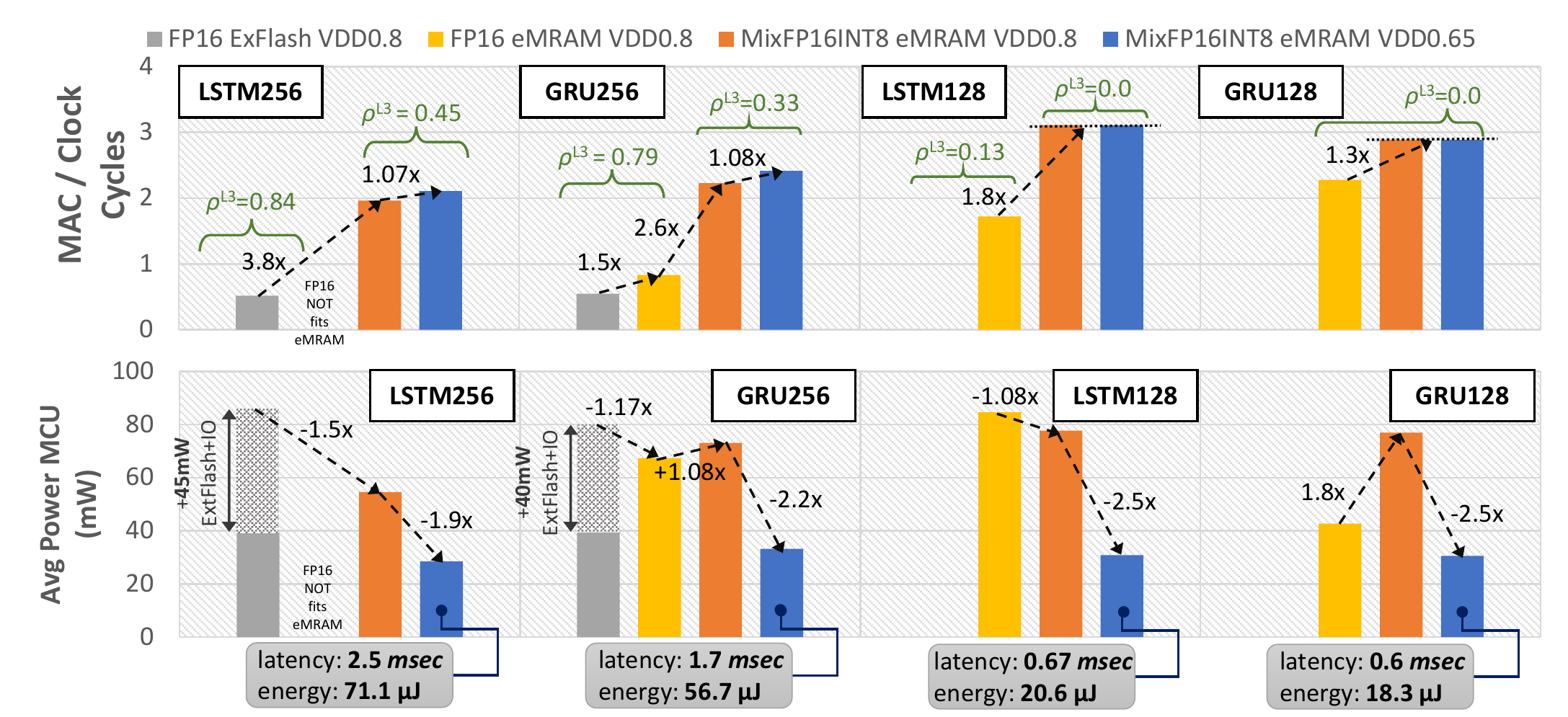}
\caption{(Top) Latency, measured in terms of MAC/cyc, and (Bottom) Power Consumption, in mW, of the TinyDenoiser models running on the target HW. } 
\label{fig:rnn-meas}
\end{figure}

\subsection{RNN-based SE inference performance on a multi-core MCU}

We analyze the effectiveness of the proposed software pipeline (Sec.~\ref{sec:swarchi}) and the novel quantization strategy by measuring the TinyDenoiser inference latency and energy consumption on a 22nm prototype of the target architecture (Sec.~\ref{sec:HWtarget}). The chip prototype can be powered at 0.8V or 0.65V with a maximum clock frequency of 370MHz and 240MHz, respectively. 
More in details, we deploy \textit{LSTM256}, \textit{GRU256}, \textit{LSTM128},  \textit{GRU128} models after \textit{FP16} and \textit{MixFP16-INT8} quantization. If exceeding 2MB of storage requirement for model parameters, we make use of an external FLASH memory while, on the contrary, the on-chip eMRAM memory can be used. This latter features a peak BW of $640 MB/sec$, independently of the voltage supply.

Figure~\ref{fig:rnn-meas} reports on the top the measured inference latencies, expressed in terms MAC/cyc, and the MCU power consumption (in mW) on the bottom. In case of FP16 \textit{LSTM256} and \textit{GRU256} models, the ratio of parameters stored in the L3 memory over the total amount of parameters, denoted as $\rho^{L3}$, achieves 0.84 and 0.79, thanks to the tensor promotion mechanism. However, the execution is L3 memory-bounded in this scenario.
In accordance to the model of Eq.~\ref{eq:latency},
the read time of a  FP16 parameter from the ExtFlash takes 2 clock cycles that explains a latency close to 0.5MAC/cyc (every MAC requires one parameter to be loaded). Because of the activity of the external memory, an extra average power cost of 40-45 mW is measured, corresponding to $\sim 50\%$ of the total power. 

While FP16 LSTM256 cannot fit the on-chip non-volatile memory, the FP16 GRU256 can cut the extra power cost by storing the FP16 parameters into the eMRAM. The MCU power consumption increases because of the on-chip memory utilization, which was OFF before, and the higher density of operations (higher MAC/cyc) due to the higher eMRAM memory BW than the ExtFlash. 

If leveraging MixFP16-INT8 for LSTM256 and GRU256, the ratio $\rho^{L3}$ decreases to 0.45 and 0.33, meaning more tensors are promoted to L2 in contrast to FP16 quantization. Thanks to this and the faster INT8 kernels, the computation efficiency increases up to 1.9 and 2.2 MAC/cyc (one of the two RNN layer is still L3 memory-bound). At the same time, the power cost of the MCU increases because of the higher operation density. Lastly, we obtain a power saving of $\sim 2\times$ by reducing the power supply to 0.65V. Also note the MAC/cyc improves by up to 8\% because the eMRAM bandwidth is independent from the system clock frequency, bringing benefits to the memory-bounded layers.

On the other side, FP16 \textit{LSTM128} and \textit{GRU128} fits the eMRAM memory capacity and show a $\rho^{L3}$ ratio as low as 0.13 and 0.0, meaning that the majority or all the memory parameters are promoted to the L2 memory before the inference. This explains the high FP16 latency efficiency, reaching up to 2.2 MAC/cyc. The MixFP16-INT8 quantization further decreases latency by $1.8\times$ and $1.3\times$.
In case of $LSTM128$ the power consumption of MixFP16-INT8 slightly decreases with respect to FP16 because eMRAM is not used, while $GRU128$ presents a $1.8\times$ higher power, in line with other settings. 
Scaling down the supply voltage do not contribute to a higher MAC/cyc metric because of the low (or null) L3 memory utilization, while the power consumption is reduced by $2.5\times$. 

Fig.~\ref{fig:rnn-meas} also reports on the bottom the latency and the energy measures for the inference tasks in the most energy efficient configuration. Even if reducing the clock frequency, the real-time constraints (6.25 msec) are matched. When considering a duty cycled operation with a sleep power much lower than the active power, the average power reduces up to 3mW for the smallest model.

\begin{table}[t]
\caption{Comparison with other SE solutions for MCUs.}
\label{tab:comparison}
\centering
\scriptsize
\begin{tabular}{|l|c|c|c|c|c|c|c|c|}
\hline
\multirow{2}{*}{\textbf{Model}} & \multirow{2}{*}{\textbf{Mpar}} & \multirow{2}{*}{\textbf{Quant}}                                               & \multirow{2}{*}{\textbf{QAT}} & \multirow{2}{*}{\textbf{Device}}                                               & \multirow{2}{*}{\textbf{Deployment}}                                              & \multirow{2}{*}{\textbf{\begin{tabular}[c]{@{}c@{}}msec/\\ inf\end{tabular}}} & \multirow{2}{*}{\textbf{\begin{tabular}[c]{@{}c@{}}MAC/\\ cyc\end{tabular}}} & \multirow{2}{*}{\textbf{\begin{tabular}[c]{@{}c@{}}MMAC/\\ W\end{tabular}}} \\
                                &                                &                                                                               &                               &                                                                                &                                                                                   &                                                                               &                                                                              &                                                                             \\ \hline
TinyLSTM~\cite{fedorov2020tinylstms}                        & 0.33                           & \multirow{2}{*}{INT8}                                                         & \multirow{2}{*}{yes}          & \multirow{2}{*}{STM32F746VE}                                                   & \multirow{2}{*}{N/A}                                                              & 4.26                                                                          & \multirow{2}{*}{0.36}                                                        & \multirow{2}{*}{0.14}                                                       \\ \cline{1-2} \cline{7-7}
TinyLSTM~\cite{fedorov2020tinylstms}             & 0.46                           &                                                                               &                               &                                                                                &                                                                                   & 2.39                                                                          &                                                                              &                                                                             \\ \hline
\multirow{2}{*}{RNNoise~\cite{jianjia_ma_2020_4158710}}        & \multirow{2}{*}{0.21}          & \multirow{2}{*}{INT8}                                                         & \multirow{2}{*}{yes}          & \multirow{2}{*}{STM32L476}                                                     & \multirow{2}{*}{\begin{tabular}[c]{@{}c@{}}NNoM w/\\       CMSIS-NN\end{tabular}} & \multirow{2}{*}{3.28}                                                         & \multirow{2}{*}{0.45}                                                        & \multirow{2}{*}{1.84}                                                       \\
                                &                                &                                                                               &                               &                                                                                &                                                                                   &                                                                               &                                                                              &                                                                             \\ \hline
\textbf{LSTM256} [ours]             & 1.24                           & \multirow{2}{*}{\begin{tabular}[c]{@{}c@{}}MixFP16\\      -INT8\end{tabular}} & \multirow{2}{*}{no}           & \multirow{2}{*}{\begin{tabular}[c]{@{}c@{}}8-core \\      RISC-V\end{tabular}} & \multirow{2}{*}{GAPFlow}                                                          & 2.50                                                                          & \textbf{2.11}                                                                & \textbf{17.78}                                                              \\ \cline{1-2} \cline{7-9} 
\textbf{GRU256} [ours]             & 0.98                           &                                                                               &                               &                                                                                &                                                                                   & 1.70                                                                          & \textbf{2.41}                                                                & \textbf{17.46}                                                              \\ \hline
\end{tabular}

\end{table}

\subsection{Comparison with other works}
Table~\ref{tab:comparison} compares our solution with state-of-the-art SE solutions on MCUs: 
TinyLSTM~\cite{fedorov2020tinylstms}, which is benchmarked on a STM32F7 MCU, and RNNoise~\cite{valin2018hybrid} deployed on a low-power STM32L4 using the NNoM software with CMSIS-NN backend~\cite{jianjia_ma_2020_4158710}.
Both solutions leverage on single-core devices and 8-bit quantization, which results effective thanks to QAT and the model design constraint of using intermediate activation features with limited numerical ranges.
Despite our solution being more subject to memory bottleneck issues because of $2.6-6\times$ more parameters and the higher bit precision, we achieve a top latency efficiency, up to $5.3\times$ and $6.7\times$ MAC/cyc higher than RNNoise and TinyLSTM, respectively. This acceleration is obtained thanks to the optimized software pipeline that efficiently exploit the underlying hardware. Additionally, the energy efficiency results up to $9.7\times$ and $123\times$ higher than previous solutions. We also remark that our solution achieves low-degradation with respect to full-precision model without relying on any expensive QAT training procedures.

\section{Conclusion}
This work proposed a novel design approach to efficiently bring RNN-based SE models on low-power multi-core MCUs. On the one side, we proposed a novel quantization scheme that mixes FP16 and INT8 PTQ to obtain low-accuracy degradation without relying on expensive QAT. On the other side, we designed an optimized software pipeline to efficiently exploit the compute performance of low-power 8-core MCU. Our design demonstrated the fastest RNN-based SE solution for MCUs, featuring $>10\times$ energy-efficiency than previous solutions.

\bibliographystyle{splncs04}
\bibliography{tinydenoiser}

\end{document}